\documentclass{article}
\usepackage{arxiv}
\usepackage[utf8]{inputenc} % allow utf-8 input
\usepackage[T1]{fontenc}    % use 8-bit T1 fonts
\usepackage{hyperref}       % hyperlinks
\usepackage{url}            % simple URL typesetting
\usepackage{booktabs}       % professional-quality tables
\usepackage{amsfonts}       % blackboard math symbols
\usepackage{nicefrac}       % compact symbols for 1/2, etc.
\usepackage{microtype}      % microtypography
\usepackage{lipsum}		% Can be removed after putting your text content
\usepackage{graphicx}
\usepackage{natbib}
\usepackage{doi}

\title{Nuclear spin self compensation system for moving MEG sensing with optical pumped atomic spin co-magnetometer}

\date{January 12, 2022}	% Here you can change the date presented in the paper title
\author{ \href{https://orcid.org/0000-0002-3169-9577}{\includegraphics[scale=0.06]{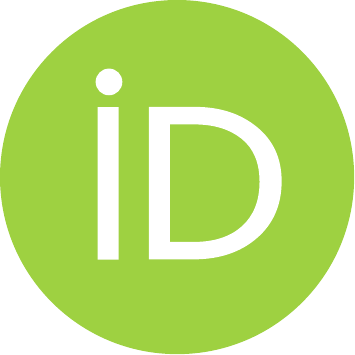}\hspace{1mm}Yao Chen$^{1,2}$, Yintao Ma$^1$, Mingzhi Yu$^1$, Yanbin Wang$^1$, Ning Zhang$^3$, Libo Zhao$^1$, and Zhuangde Jiang$^1$}\\
	1. School of Mechanical Engineering,\\
	State Key Laboratory for Manufacturing Systems Engineering,\\ International Joint Laboratory for Micro/Nano Manufacturing and Measurement Technologies,\\ Overseas Expertise Introduction Center for Micro/Nano Manufacturing-\\
	and Nano Measurement Technologies Discipline Innovation,\\ Xi'an Jiaotong University, Xi'an 710049, China\\
    2.Xi'an Jiaotong University Suzhou Institute, Suzhou 215123, China.\\
    3.Research Center for Quantum Sensing, Intelligent Perception Research Institute, \\Zhejiang Lab, Hangzhou 310000, China.\\
		\texttt{libozhao@xjtu.edu.cn} \\
		\texttt{yaochen@xjtu.edu.cn} 
}

\renewcommand{\shorttitle}{\textit{arXiv} Template}

\hypersetup{
pdftitle={A template for the arxiv style},
pdfsubject={q-bio.NC, q-bio.QM},
pdfauthor={Yao Chen, Yintao Ma, Mingzhi Yu, Yanbin Wang, Zhuangde Jiang, Libo Zhao},
pdfkeywords={Atomic Magnetometer, Magnetoencephalography, Atomic Co-magnetometer, Atomic Spin Gyroscope, Atomic Magnetometer, Magnetoencephagraphy},
}

\begin{document}
\maketitle

\begin{abstract}
Recording the moving MEGs of a person in which a person's head could move freely as we record the brain's magnetic field is a hot topic in recent years. It is well known that the fluctuation background magnetic field is the main noise source for MEGs' recording as the head is moving and the key is to develop method for background magnetic field compensation.  Traditionally, atomic magnetometers are utilized for moving MEGs recording and a large compensation coil system is utilized for background magnetic field compensation. Here we described a new potential candidate: an optically pumped atomic co-magnetometer(OPACM) for moving MEGs recording. In the OPACM, hyper-polarized nuclear spins could produce a magnetic field which will shield the background fluctuation low frequency magnetic field noise while the the fast changing MEGs signal could be recorded. The nuclear spins look like an automatic magnetic field shields and dynamically compensate the fluctuated background magnetic field noise. In this article, the magnetic field compensation is studied theoretically and we find that the compensation is closely related to several parameters such as the electron spin magnetic field, the nuclear spin magnetic field and the holding magnetic field. Based on the model, the magnetic field compensation could be optimized. We also experimentally studied the magnetic field compensation and the responses of the OPACM to different frequencies of magnetic field are measured. We show that the OPACM owns a clear suppression of low frequency magnetic field under 1Hz and response to magnetic field's frequencies around the band of the MEGs. Magnetic field sensitivity of  $3fT/Hz^{1/2}$ has been achieved. Finally, we do a simulation for the OPACM as it is utilized for moving MEGs recording. For comparison, the traditional compensation system for moving MEGs recording is based on a coil which is around 2m in dimension while our compensation system is only 2mm in dimension. Moreover, our compensation system could work in situ and will not affect each other.
\end{abstract}

% keywords can be removed

\section{Introduction}
Magnetoencephalography(MEG) finds wide application in several diseases including diagnosing early stages of Alzheimers' disease\cite{megalzheimer}, localization of the epileptogenic zone before epilepsy surgery\cite{MEG2015epileptogenic}, studying of depression\cite{MEGDEPRESSION}, et al. With the development of the spin exchange relaxation free(SERF) atomic magnetometer\cite{kominis2003subfemtotesla} which could work under room temperature, it is a competitor for SQUID(super conducting quantum interference devices) magnetometers which is a typical equipment for MEGs recording. Shortly after the invention of the SERF atomic magnetometer, it is utilized for MEG study and the auditory evoked magnetic field is recorded\cite{xia2006magnetoencephalography,Colombo:16,Borna_2017}. 

Though there are a lot of studies about utilizing atomic magnetometer for acquiring MEGs including auditory evoked brain magnetic feild\cite{xia2006magnetoencephalography,shah2013compact},somatosensory evoked magnetic fields\cite{borna2020non,ucl2019sensory},visual evoked magnetic fields\cite{labyt2018magnetoencephalography},etc. rare studies have been carried out to study the motor system while the subject is free to move\cite{boto2018}. The motor system include naturally walking of a man as the MEG is recorded, head moving muscular related MEGs in the motor cortex, neurological disorders induced essential tremors, the MEGs of visual-motor integration, etc. 

Several studies have been carried out to study the moving MEGs. We know that the brain's magnetic field is so small compared with the earth's magnetic field that the current SERF magnetometers need to work in a magnetic shield room for MEGs recording and the residual magnetic field need to be under 2nT\cite{holmes2018bi}. Not only the residual magnetic fields in the shield needs to be compensated at three directions as well as their gradients need to be compensated. If there are magnetic field gradients, the magnetic field experienced by the sensors will change as the sensor is moving. Moreover, fluctuation from the environment such as the nearby subway trains will also affect the magnetometers. The key for moving MEGs recording is to develop an active compensation system to suppress the background magnetic field. 

In the method developed by E. Boto et al.,\cite{boto2018} for moving MEGs recording, a 1.6m$\times$1.6m bi-planar coil\cite{holmes2018bi} is utilized for nulling the background magnetic field as well as its gradient. The wearable magnetometer system only can work in space of $40cm\times40cm\times40cm$\cite{boto2018}. Feedback system is developed to maintain the residual magnetic field in this area to be below 2nT in real time. Moreover, the person's head only can move slowly and in a small area as the MEGs are recorded. In the method developed by S. Mellor et al.\cite{mellor2021magnetic}, the spatial variation in the magnetic field is modelled and the model is used to predict the movement related artefact magnetic field\cite{rea2021precision,tierney2021modelling}. Real time brain magnetic field could be extracted by subtracting the artefact magnetic field from the magnetometer readings. Note that this method depends on predicting the magnetic field inside the magnetic field shield room. It is only effective for stable magnetic field. For changing background magnetic field, for example the magnetic field originate from the subway train, its performance should be worse. In the method developed by Kernel Flux\cite{pratt2021kernel}, a whole-head 432-magnetometer optically-pumped
MEG system is developed and it is allowed for comfortable head motion. In order to achieve head motion, the system’s common mode rejection ratio (CMRR) owns the ability to distinguish nearby signal sources from distant noises such as the background fluctuation magnetic field noise. Since this method utilize the differential technique and only the gradient of the brain's magnetic field could be measured. It could not directly measure the brain's magnetic field.

This article describes a new method to do this kind of background magnetic field compensation. Different from the traditional atomic magnetometers, we use an OPACM to do moving MEGs measurement. In the OPACM, nuclear spins are filled in the vapor cell and the hyper-polarized nuclear spins produce a magnetic field which will automatically shield the electron spins in the OPACM from the environment fluctuation magnetic field. This will leave the electron spins only sensitive to the higher frequency MEG signal. In our method, we don't need a large compensation coil system and the feedback system which owns a dimension of $1.6m\times1.6m$. Our compensation system works in situ and only owns a dimension around $2mm\times2mm\times2mm$. A person could move more naturally in the magnetic field shielding room and this would give the change to record MEGs as a person is walking naturally. Moreover, our method record MEGs directly without differential of the background noise. Our method is also effective to background magnetic field noise such as noise from the subway trains.

The automatically compensated process  process is similar to the self compensation effect in an SERF co-magnetometer\cite{kornack2005,yao2016}. Due to its self compensating ability for the DC magnetic field, SERF atomic co-magnetometers have been utilized for rotation sensing\cite{li2016rotation,duan2018rotation} or searching physics beyond the standard model\cite{justingbrown2010,jiwei2018,vasilakis2009limits}. For rotation sensing, the atomic spins are directed in the inertial space and its direction could be easily disturbed by the background magnetic field fluctuation. Thus the self compensating effect could greatly reduce the co-magnetometer's sensitivity to magnetic field. For detection the physics beyond the standard model, similar situation happens. 

Note that atomic magnetometers fabricated by micro-machining technology is also fast developing and the size of the magnetometer is greatly reduced. It is a fascinating research trend and we can find some good works in these references\cite{Sander:12,Alem:17}. Not only MEGs are studied by the atomic magnetometer, but also fetal magnetocardiography measurements were also studied\cite{Alem_2015}. We believe that our OPACM also could be fabricated by the micro machining technology and we are now developing OPACMs by MEMS technology.

\section{Theory}
\label{sec:theory}
\begin{figure}
	\centering
\includegraphics[width=13cm,height=8cm]{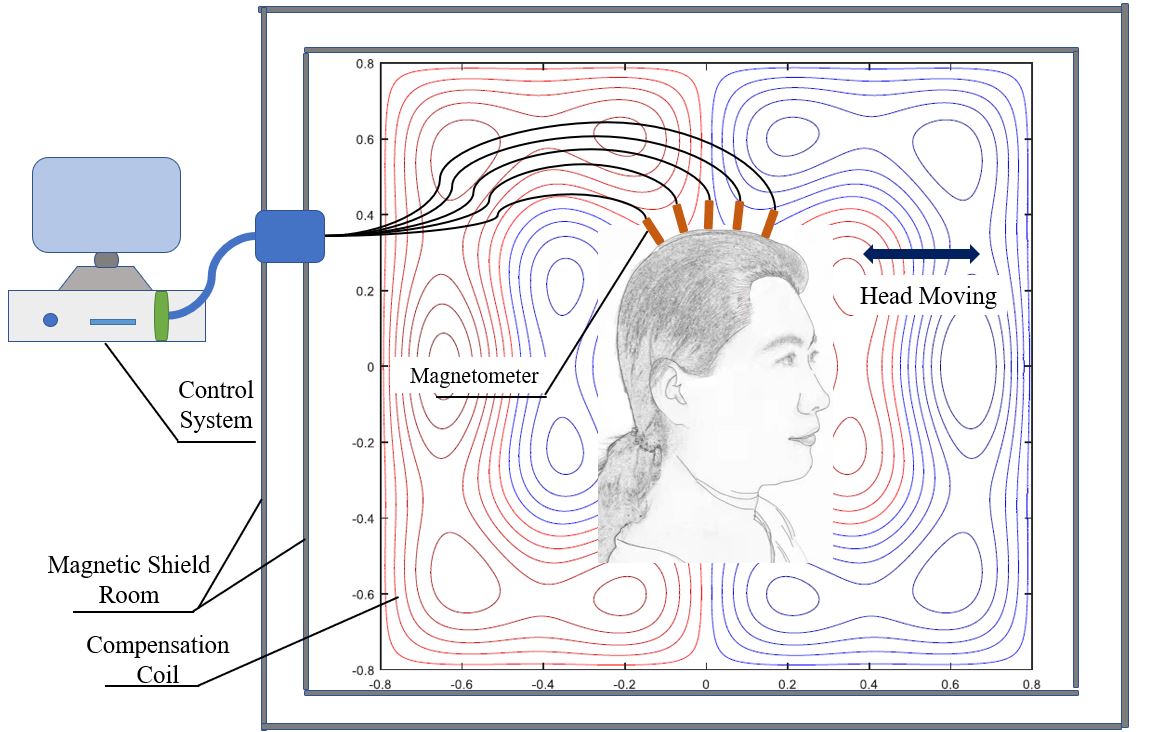}
	\caption{The picture illustrates the brain magnetic field measurement as the head is moving. The person who need to do MEG recording stays in a magnetic field shielding room. Further magnetic field compensation is done by the planar coil. As the head is moving, due to the inhomogeneous of the background field. The sensors will experience fluctuation magnetic field around 1nT typically. This field strength is much larger than the brain magnetic field which is around 100fT. Thus, we have developed an automatic compensation system through hyper-polarized nuclear spins. This nuclear spin shielding will leave the magnetometer be sensitive to the fast changing brain magnetic field. The brain magnetic field will be recorded by the sensors and the control system will analysis the signal.}
	\label{fig:1}
\end{figure}

\begin{figure}
	\centering
\includegraphics[width=16cm,height=9cm]{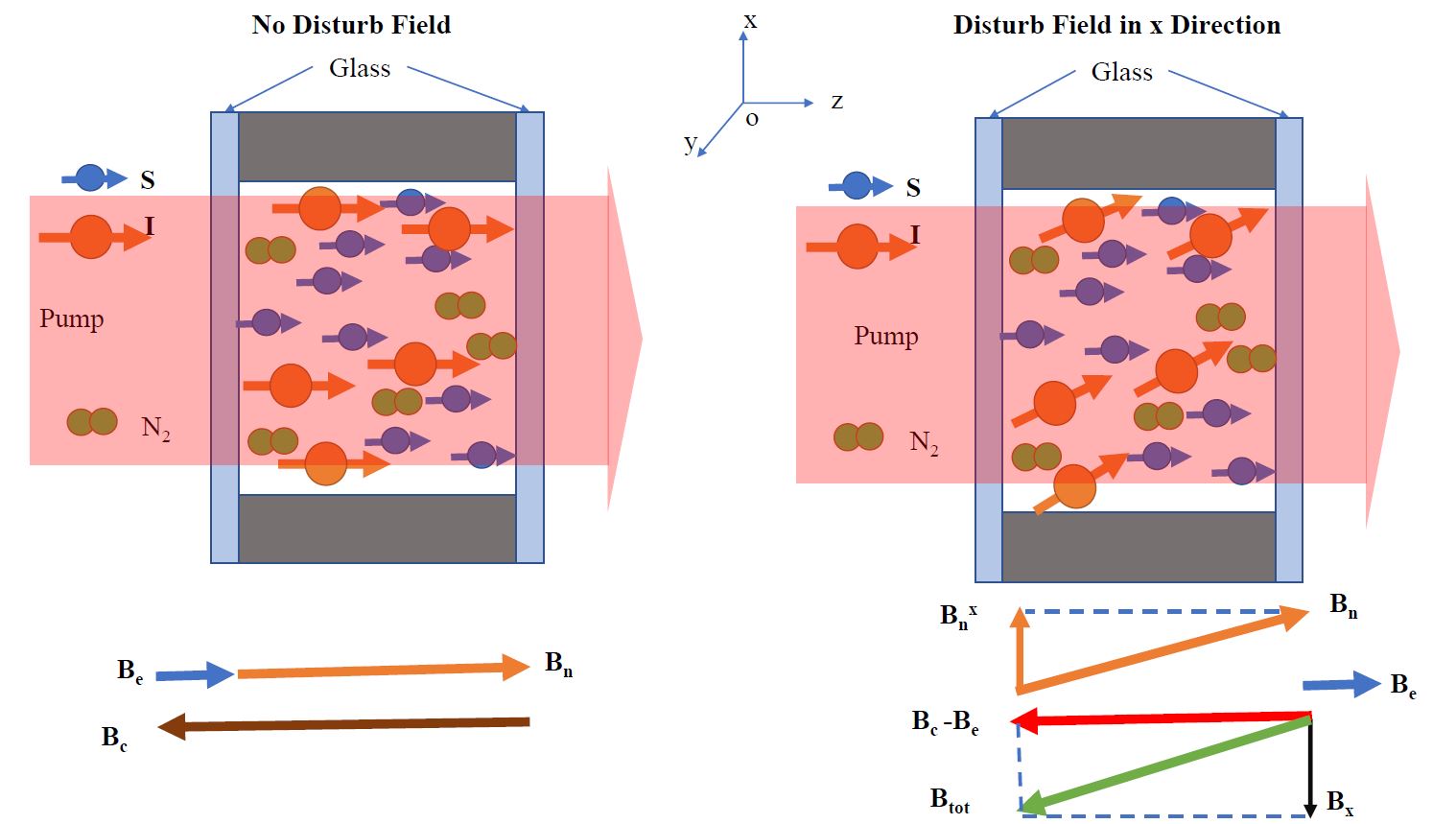}
	\caption{The principle of the fluctuating background magnetic field compensation.}
	\label{fig:2}
\end{figure}

 As shown in Figure.\ref{fig:1}, due to the ultra low intensity of the brain's magnetic field, the measurement of the MEGs are done in a magnetic field shielding room with the inner space to be typically around $2m\times2m\times2m$\cite{altarev2014magnetically}. Further magnetic field compensation is carried out by the coils inside the room. Usually the residual magnetic field around the head area which is typically $0.5m\times0.5m\times0.5m$\cite{holmes2018bi} could be reduced to under 1nT with active compensation of the background magnetic field and its gradient by the coils. However, this is still much larger than the brain magnetic field. Due to the gradient of the residual magnetic field, there will be substantial variation as the head moves in the room. The typical SERF atomic magnetometer is sensitive to both of the background fluctuation magnetic field and the brain magnetic field. We notice that the background magnetic field and its fluctuation are typically low changing magnetic field while the brain magnetic field is fast changing field. If we can automatically compensate the background low frequency magnetic field and leave the magnetometer sensitive to the fast changing brain magnetic field. That will allow the brain's magnetic field to be recorded as the head is moving in the magnetic shield room. \\
The key to measure brain magnetic field is the SERF magnetometer. While the reason why that the SERF magnetometer owns very high sensitivity is caused by the reducing of the spin exchange relaxation between electron spins collision. The idea of spin exchange relaxation suppression was first developed by William Happer's group\cite{happer1973spin}. Not until 2002 the spin exchange relaxation free atomic magnetometer was invented\cite{allred2002high} and its sensitivity surpass that of a SQUID magnetometer\cite{kominis2003subfemtotesla}. In a SERF atomic magnetometer, the electron spins of the alkali atoms are optically pumped by a laser and then the spins will rotate an angle if magnetic field is experienced by the spins. Both of the optical rotation or the laser absorption method could be utilized for detection of the polarized spins' direction. Thus the magnetic field could be indirectly deduced by the optical detection method.\\
If we fill nuclear spin $I$ in the vapor cell, under certain conditions, these nuclear spins will shield the electron spins from the outside fluctuation magnetic field. As shown in Figure.\ref{fig:2}, nuclear spins  with momentum $I$ and electron spins with momentum $S$ are filled in the vapor cell. The electron spin $S$ are optically pumped by the laser light and point to the direction of the laser light. Then the nuclear spin $I$ will be hyper-polarized by the electron spins and there polarization direction is the same as that of the electron spins\cite{walker1997}. Due to the Fermi contact interaction\cite{walker1989}, the hyper-polarized nuclear spins will produce magnetic field which could be experienced by the electron spins. In a spherical vapor cell, the effective magnetic field will be approximately $B^n=8\pi k_0 \mu_n [N] P^n/3$\cite{romalis1998}. In the equation, $B_n$ is the magnetic field produced by the nuclear spins such as $^{21}Ne$ or $^{131}Xe$. $k_0$ is an enhancement factor\cite{walker1989} which could enhance the magnetic field experienced by the electron spins during spin exchange collision. $\mu_n$ is the nuclear magnetic moment for nuclear spin $I$. $[N]$ is the number density of the nuclear spins and $P^n$ is the polarization of the nuclear spins. Note that the $^{21}Ne$ nuclear magnetic moment  $\mu(^{21}Ne)$ is 0.66$\mu_N$ in which $\mu_N$ is the nuclear magnetic moment of the neutron and $^{131}Xe$ nuclear magnetic moment $\mu(^{131}Xe)$ is 0.69$\mu_N$. Just as the nuclear spins, the electron spins will also produce a magnetic field $B_e$ which will be experienced by the nuclear spins and $B_e$ is equal to $8\pi k_0 \mu_B [A] P^e/3$. Here $\mu_B$ is the Bohr magneton, $[A]$ is the number density of the electron spins and $P^e$ is the polarization of the electron spins. Under steady state, the polarization of the nuclear spins and electron spins will be stable and $B_n$ and $B_e$ is stable. In order to realize the automatic compensation effect, we add a compensation magnetic field $B_c$ in the opposite direction of $B_n$ and the strengths of $B_c$ is equal to the strength of $B_n$+$B_e$.\\
As shown in Figure.\ref{fig:2}, if there is a disturb background magnetic field $B_x$, the total magnetic field $\mathbf{B_{tot}}$ experienced by the nuclear spins will be $\mathbf{(B_c-B_e)+B_x}$. Since $B_c-B_e$ is equal to $B_n$, under the small disturb magnetic field which means that $B_x$ is much smaller than $B_n$, the projection of $\mathbf{B_n}$ along the $x$ direction $B_n^x$ is equal to $B_x$ and they are in the opposite direction. Thus, the electron spins will still stay in the pumping laser direction. We say that the nuclear spins automatically compensate the fluctuation magnetic field and shield the electron spins from the disturbing magnetic field.\\
Note that the nuclear spins can perfectly shield the DC disturbing magnetic field theoretically while the AC magnetic field from the brain, etc. could not be compensated perfectly by the nuclear spins because the finite bandwidth of the nuclear spins' response to magnetic field. This will give the chance of the magnetometer only sensitive to the fast changing brain magnetic field while automatically compensate the background low frequency disturbing magnetic field. To further study this mechanism in details, we need to solve the Bloch equations of this system and the response of the magnetometer to several frequencies of magnetic fields need to be studied.\\
In this paper, we mainly focus on the magnetometer based on K-Rb-$^{21}$Ne. The electron spins are mainly from the Rb atoms. Since K atoms are utilized as the hybrid optical pumping atoms, typically K atom spins account for only very small part of the electron spins. Most of the electron spins are from the Rb atoms\cite{yao2016collisionmixing}. We can neglect the electron spins from the K atoms. The full Bloch equations could be simplified to just include two spin species. There are various literature describe the full Bloch equations for the K-Rb-$^{21}$Ne magnetometer\cite{yao2016,kornack2002dynamics}. Here we briefly list them.
\begin{equation} \label{Equation.1}
\frac{\partial \mathbf{P^e}}{\partial t}=\frac{\gamma^e}{Q(P^e)}(\mathbf{B}+\lambda_{Rb-Ne}M_0^n\mathbf{P^n})\times \mathbf{P^e}-\frac{1}{T_{2e},T_{2e},T_{1e}}\mathbf{P^e}/Q(P^e)+(R_p\mathbf{s_p}-R_p\mathbf{P^e})/Q(P^e)
\end{equation}
\begin{equation} \label{Equation.2}
\frac{\partial \mathbf{P^n}}{\partial t}=\gamma^n(\mathbf{\Omega}/\gamma^n+\mathbf{B}+\lambda_{Rb-Ne}M_0^e\mathbf{P^e})\times \mathbf{P^n}+R_{Rb-Ne}^{se}(\mathbf{P^e}-\mathbf{P^n})-\frac{1}{T_{2n},T_{2n},T_{1n}}\mathbf{P^n}
\end{equation}
Here $\mathbf{P^e}$ and $\mathbf{P^n}$ are the Rb electron and $^{21}$Ne nuclear spin polarizations, $\mathbf{\Omega}=\{ \Omega_x,\Omega_y,\Omega_z\}$ is the rotation angular velocity input, $\mathbf{B}$ is the external magnetic field which include the compensation field $\mathbf{B_c}$ and the disturbing field $\mathbf{B_x}$, $M_0^e=\mu_B [A]$ and $M_0^n=\mu_N[N]$ are the magnetizations of the electron and nuclear spins as the spins are fully polarized,$\lambda_{Rb-Ne}$ is equal to $8\pi k_0/3$ for the spherical alkali vapor cell, $R_p$ is an effective pumping rate which is related to the K pumping rate and the density ratio of K to Rb\cite{yao2016collisionmixing}, $\mathbf{s_p}$ is the optical pumping vector along the propagation of the pump with magnitude equal to the degree of circular polarization, $R_{Rb-Ne}^{se}$ is the spin exchange rate of $^{21}$Ne nuclear spins. The precession frequency of the alkali metal is slowed by factor $Q(P^e)$\cite{Savukov:2005} which is related to the alkali spin polarization. $T_1$ and $T_2$ are the relaxation times for components of the polarization parallel and transverse to $\mathbf{B_z}$, respectively. The subscipts $e$ and $n$ denote for the electron spins and nuclear spins.\\
Here we are interested in the magnetometer's sensitivity to external fluctuation magnetic field. Thus we need to study the magnetometer's responses to external magnetic field under different frequencies. A sinusoidal signal will be applied to the Bloch equations to get the output amplitude. \\ 
It is reasonable to assume that the disturbing field is much smaller than the compensation field, thus the polarization of the electron spin and the nuclear spin in the $z$ direction is constant. We could throw away the $P_z^e$ and $P_z^n$ in Equation(\ref{Equation.1}) and Equation(\ref{Equation.2}). We define $\mathbf{P}=\{ P_x^e,P_y^e,P_x^n,P_y^n \}^T$ and Equation(\ref{Equation.1}) and Equation(\ref{Equation.2}) could be simplified to be $d\mathbf{P}/dt=\mathbf{A} \bullet \mathbf{P}+\mathbf{C}$. Since we need nearly 5 hours to polarize $^{21}$Ne. Thus, it is reasonable to assume that $R_{Rb-Ne}^{se}$ and the relaxation rate of $^{21}$Ne are small numbers. The equations could be simplified and $\mathbf{A}$ is equal to:
\begin{equation}\label{Equation.3}
\left(
\begin{array} {cccc}
\tilde{-R_{tot}^e} & -\omega_e & 0 & \omega_{en}\\
\omega_e & \tilde{-R_{tot}^e} & -\omega_{en} & 0\\
0&\omega_{ne} & 0& \omega_n\\
-\omega_{ne} & 0& \omega_n&0\\
\end{array} 
\right)
\end{equation}
In the above equation, $\tilde{-R_{tot}^e}$ is equal to $(1/T_{2e}+R_p)/Q(P^e)$, $\omega_{en}$ is equal to $\gamma^e \lambda_{Rb-Ne}M_0^n {P_z^e}/Q(P^e)$, $\omega_{ne}$ is equal to $\gamma^n \lambda_{Rb-Ne}M_0^e {P_z^n}$, $\omega_n$ is equal to $\gamma^n(B_c-B_e)$. The input term $\mathbf{C}$ is equal to:
\begin{equation}\label{Equation.4}
\{ \tilde{b_y^e}=P_z^e \gamma^e B_y/Q(P^e),\tilde{-b_x^e}=-P_z^e \gamma^e B_x/Q(P^e),\tilde{b_y^n}=P_z^n \gamma^n B_y,\tilde{-b_x^n}=-P_z^n \gamma^n B_x \}^T
\end{equation}
Suppose that there are two oscillating magnetic field inputs in the $x$ and $y$ direction:
\begin{equation}\label{Equation.5}
\mathbf{B}=(B_{0x}\mathbf{x}+B_{0y}\mathbf{y})e^{-i\omega t}
\end{equation}
Under steady state, the polarization of the electron spin and the nuclear spin could be solved:
\begin{equation}\label{Equation.6}
\mathbf{P}=(\mathbf{A}-i\omega \mathbf{I})^{-1}\bullet\mathbf{C}
\end{equation}
In the experiment, the $x$ polarization of the electron spins are detected through a probe light. Under the typical conditions, the electron precession frequency $\omega_e$ is much larger than that of the nuclear spin frequency $\omega_n$ and the input magnetic field frequency $\omega$. Equation(\ref{Equation.6}) could be further simplified to be:
\begin{equation}\label{Equation.7}
P_x^e\approx\frac{B_{0x}\gamma^eP_z^e\omega(\omega\omega_e-i\tilde{R_{tot}^e}\omega_n)e^{-i\omega t}}{Q(P^e)\left[\tilde{R_{tot}^e}(\omega-\omega_n)+i\omega\omega_e \right]\left[ \tilde{R_{tot}^e}(\omega+\omega_n)-i\omega\omega_e\right]}
\end{equation}
Equation(\ref{Equation.7}) contains both of the real part and the image part. The real part represent the amplitude of the output signal while the image part represent the phase shift. Thus we just consider the real part to deduce the response of the magnetometer to external magnetic field. Note that Equation\ref{Equation.7} only consider the $x$ oscillating magnetic field response. If we consider both of the $x$ and the $y$ direction, the equation could be further simplified:
\begin{equation}\label{Equation.8}
P_x^e\approx P_z^e\frac{\gamma^e B_{0x}\omega e^{-i(\omega t+\Phi_x)}}{R_{tot}^e\left[\omega_n^2+\omega^2\omega_e^2/(R_{tot}^e/\gamma^e)^2\right]^{1/2}}+P_z^e\frac{\gamma^e B_{0y}\omega^2 e^{-i(\omega t+\Phi_y)}}{R_{tot}^e\left[\omega_n^2+\omega^2\omega_e^2/(R_{tot}^e/\gamma^e)^2\right]}
\end{equation}
$\Phi_x$ and $\Phi_y$ are the phase shifts between the input oscillating magnetic field and output signal for the $x$ and $y$ direction respectively. From Equation(\ref{Equation.8}) we can see that the output signal is frequency dependent and related to $\omega_n$,$\omega_e$ and $R_{tot}^e$. Under low frequency, the output signal is proportional to $\omega$ in the $x$ direction and $\omega^2$ in the $y$ direction. The lower the frequency, the smaller the output signal. That means the magnetometer could efficiently suppress the low frequency disturbing background magnetic field while leave it sensitive to the higher frequency brain magnetic field. We can also change the parameters to let the magnetometer work in a proper frequency range. Note that numerical solution could also be done based on the Equation(\ref{Equation.1}) and Equation(\ref{Equation.2}). The analytical solution in Equation(\ref{Equation.8}) has limitation. The solution assume that the the input magnetic field frequency $\omega$ is much smaller than $\omega_e$. At high frequency, the result will deviated from the solution. Under this circumstance, the numerical solution should be utilized.\\
\section{Experimental Setup}
\label{sec:experiment}
\begin{figure}
	\centering
\includegraphics[width=16cm,height=10cm]{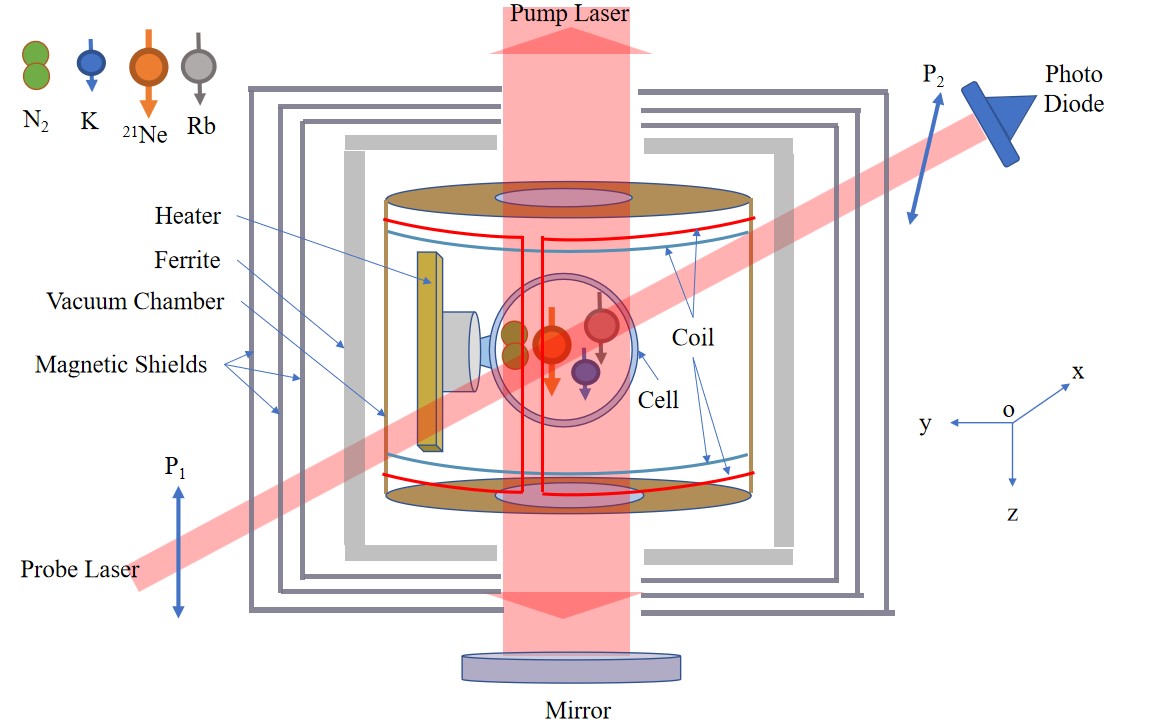}
	\caption{The schematic of the experimental setup.}
	\label{fig:3}
\end{figure}
The configuration of the magnetometer is similar to the configurations described in these references\cite{yao2016,fang2016low}. The schematic of the experimental setup is shown in Figure.\ref{fig:3}. The homemade vapor cell utilized in this experiment is spherical and the material is aluminosilicate. The diameter is 14 mm and a small droplet of K-Rb ($^{85}$Rb (72.2$\%$) and $^{87}$Rb (27.8$\%$)) mixture is contained in it. $^{21}$Ne gas (70$\%$ isotope enriched) under a pressure of about 3 Atm and N$_{2}$ gas under about 40 Torr for quenching are filled. The mole fraction ratio of K in the mixture is approximately 0.05. An homemade 110 kHz AC electrical heater is utilized for non-magnetic heating of the vapor cell. For better thermal conductivity, a nitride ceramic post was connected between the vapor cell and the heater. The heater and the vapor cell are stayed in a vacuum chamber which is made of PEEK. 0.1 Pa vacuum was achieved by using 10 L/s turbo molecular pumps for thermal insulation. Several layers of $\mu$-metal magnetic field shields together with a layer of 10-mm-thick ferrite\cite{Kornack2007} were used to shield the vapor cell from earth’s magnetic field and other fluctuation magnetic field. The diameter of the ferrite is 100 mm. Coils are utilized for further residual magnetic field compensation. In order to achieve high $^{21}$Ne polarization, the K-Rb hybrid pumping technique\cite{romalis2010hybrid} was utilized\cite{Smiciklas:2011}. K atoms are directly pumped and Rb atoms are polarized through spin exchange optical pumping with K atoms. Since the low efficiency of the hybrid optical pumping, most part of the laser light is wasted and laser with around 1W power is utilized. In order to efficiently pump the atoms, the pumping light is reflected by a mirror again after passing through the vapor cell. The projection of Rb electron spin along the probe light direction $P_x^e$ is detected by a linearly polarized light. If there is polarization projection along the probe light direction, the initial polarization plane $P_1$ will rotates an angle after passing the vapor cell and the final polarization plane will be $P_2$. The polarization rotation is detected with the Photo Elastic Modulation method\cite{yao2016}. The probe distributed feedback (DFB) laser was detuned approximately 0.4 nm away from the absorption centre of the Rb D1 line.\\

\section{Results}
\label{sec:results}
\begin{figure}
	\centering
\includegraphics[width=12cm,height=8cm]{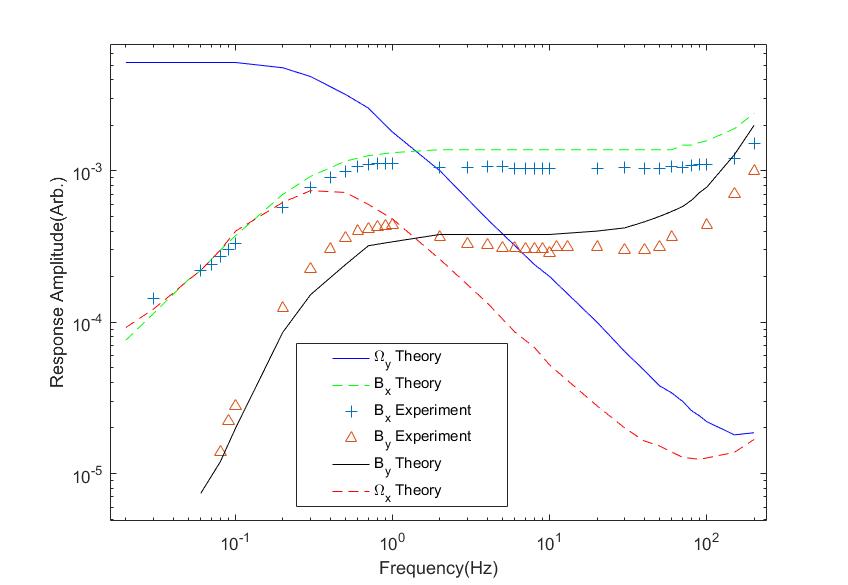}
	\caption{The responses of the magnetometer to several input signals.}
	\label{fig:4}
\end{figure}

\begin{figure}
	\centering
\includegraphics[width=14cm,height=8cm]{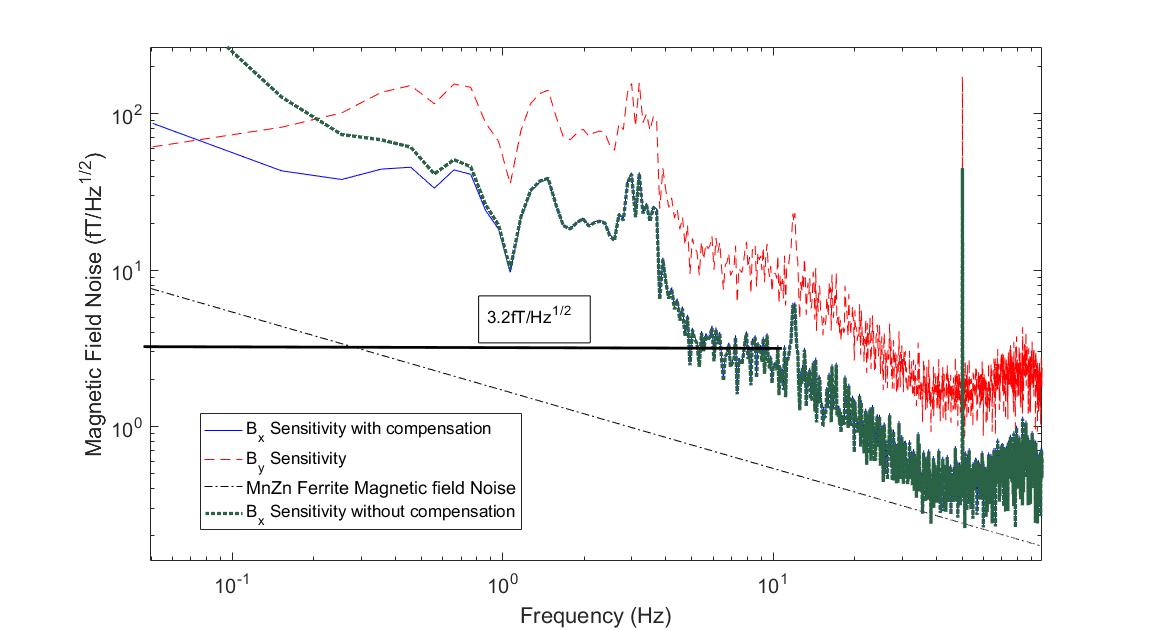}
	\caption{The sensitivity of the magnetometer.}
	\label{Figure.6}
\end{figure}

The key for a good simulation is to determine the parameters for the simulation. Here we will list them. The temperature of the vapor cell is around 473K which is measured through a temperature sensor. From the mole fraction ratio we can calculate that the number density of K and Rb are $n_K=8.1\times10^{12}/cm^3$ and $n_{Rb}=8.7\times10^{14}/cm^3$ respectively. The number density of Rb is much larger than that of the K. Since $n_{Rb}$ is very large, the electron spin will produce around $B^e$=100nT magnetic field which will be experienced by the $^{21}$Ne nuclear spins. This magnetic field is acquired as the Rb polarization is around 50\% which is the optimized polarization. Under the 50\% polarization condition, the magnetometer owns a best response. At the same time, the nuclear spins will produce a magnetic field witch is around $B^n$=500nT. The total relaxation of Rb includes the optical pumping rate, the Rb-Rb spin destruction rate, the Rb-Ne spin destruction rate and the spin exchange relaxation. Especially in the K-Rb-$^{21}$Ne magnetometer, the large $B^e$ will lead to a large magnetic field experienced by the Rb electron spins and this will lead to the spin exchange relaxation of Rb\cite{yao2016}. \\
Based on the method described in the reference\cite{yao2016}, the optical pumping rate of Rb $R_p$ is 1950s$^{-1}$ and the total relaxation rate of Rb $R_{tot}^e$ is 3900s$^{-1}$. The pump laser power density is 640mW/cm$^2$. Under this power, the Rb polarization is around 50\%. We also measured the spin destruction rate of Rb to be 480s$^{-1}$. From the total relaxation rate, we substract the optical pumping rate and the spin destruction rate and then we can get the spin exchange relaxation of Rb-Rb to be 1470s$^{-1}$. Under the temperature of our experiment, the spin exchange rate between K and Rb is $1.6\times10^5s^{-1}$. This rapid spin exchange rate will mixed K and Rb together and they seems to be like just one species\cite{yao2016collisionmixing}. K atoms are directly pumped by the laser. Through the optical pumping rate of Rb and the number density ratio of Rb to K we can calculate that the optical pumping rate for K is 210000s$^{-1}$.\\
In order to acquire the responses of the magnetometer to different frequencies of magnetic field, we theoretically applied the oscillating magnetic field into Equation(\ref{Equation.1}) and Equation(\ref{Equation.2}). With the experimental parameters, we can calculate the output response $P_x^e$. The input magnetic field amplitudes in the $x$ and $y$ direction are the same $B_{0x}=B_{0y}=0.08nT$. This magnetic field is much smaller than the equivalent magnetic field line width. Thus the co-magnetometer works in the linear area. As the co-magnetometer probes the $x$ polarization of the alkali metal spins, we directly give the amplitude of the $x$ polarization sinusoidal output signal. Figure.\ref{fig:4} shows both of the theoretical and experimental results of the $x$ and $y$ magnetic fields responses. The simulation results are done by the numerical calculation based on Equation(\ref{Equation.1}) and Equation(\ref{Equation.2}). We fit the experimental results to the simulation results. In Figure.\ref{fig:4}, '$B_x$ theory' and '$B_y$ theory' means the theoretical simulation of the response of the magnetometer to oscillating magnetic field under different frequencies. The frequency range is from 0.03Hz to 200Hz. '$B_x$ Experiment' and '$B_y$ Experiment' means the experimental results. In the experiment, we applied both of the $x$ and $y$ direction magnetic field into the magnetometer and then we recorded the output signal amplitude. The output signal's frequency is the same as the input signal's frequency. The input magnetic field amplitude is 0.08nT. From Figure.\ref{fig:4} we can see that the experiment result and the theoretical result fit well with each other. This result gives a very good evidence that our simulation model is right and the parameters in the model are reasonable. \\
It is not surprise that the frequency responses of the magnetometer to magnetic fields at frequencies under 1Hz is small. We have theoretically shown that the magnetometer is not sensitive to DC magnetic field due to the nuclear spins' self compensation mechanism. From Equation(\ref{Equation.8}), we can simplify the equations under very low $\omega$. We can get that the amplitude of the $x$ magnetic field response is proportional to $\omega/\omega_n$. As the frequency of the magnetic field reduced to 1/10 of the original frequency, the amplitude also reduced to 1/10 of the original amplitude. For the $y$ direction, the amplitude is proportional to $(\omega/\omega_n)^2$. This means that if the frequency of the magnetic field reduces to 1/10 of the original frequency, the amplitude reduces to 1/100 of the original amplitude. The magnetometer owns higher ability of magnetic field suppression in the $y$ direction. At higher frequencies, the response is larger. This is a fact that the nuclear spins and the electron spin ensembles will coupled together at low frequencies to automatically compensate the input magnetic field.\\
Since the magnetometer is sensitive to rotations\cite{yao2016,li2016rotation}, we also theoretically studied the magnetometer's responses to the input of $x$ and $y$ angular velocity input. Like magnetic field, oscillating $\Omega_x$ and $\Omega_y$ are applied to the magnetometer and the output amplitudes are recorded under different frequencies. From Equation(\ref{Equation.2}) we can see that the rotations could be equivalent to magnetic field through dividing by the gyroscopic ratio $\gamma^n$. Thus, we let the rotation angular velocity amplitudes be equal to 0.08nT which is the same as the magnetic fields' input amplitude. From Figure.\ref{Equation.4} we can see that under the low frequencies the magnetometer is sensitive to the $y$ angular velocities. If the input angular velocity frequency is much larger than 1Hz, the output amplitude signal is greatly suppressed. While for $\Omega_x$, the magnetometer is not sensitive to the $x$ rotations. Note that out experiment was done on a large and heavy platform, it is impossible to drive the platform by a rotation platform. Thus we only show the theoretical results of the rotation angular velocity responses. This could be studied in the future by a miniature magnetometer.\\ 
\begin{figure}
	\centering
\includegraphics[width=16cm,height=6cm]{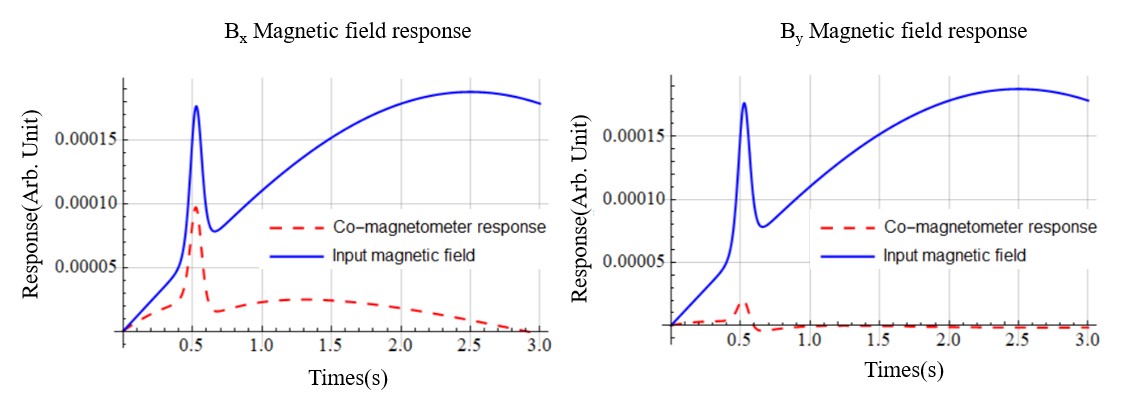}
	\caption{The magnetometer's responses to a fast pulse magnetic field and a low changing sinusoidal magnetic field in both of the $x$ and $y$ directions.}
	\label{Figure.5}
\end{figure}
In order to see if the magnetometer could suppress the low changing magnetic field while retain its sensitivity to fast change brain magnetic field. We did a theoretical response study in the time zone. As shown in Figure.\ref{Figure.5}, the input magnetic field in the $x$ and the $y$ direction is the same. The signal includes a fast changing pulse signal and a slow changing sinusoidal signal. The frequency of the low frequency sinusoidal signal is 0.1Hz which fulfill the requirement of low frequency conditions in Figure.\ref{fig:4}. For the brain magnetic field, we assume that the width of the pulse is under 0.25s which is reasonable since it is reported that the width of the auditory evoked brain magnetic field is around 0.25s\cite{xia2006magnetoencephalography}. The dashed lines show the responses of the magnetometer. For a better comparison, we let the peak of the pulse is the same as the peak of the low frequency sinusoidal signal. We can see a clear peak and low frequency magnetic field suppressed output signal. For the $x$ magnetic field response, the peak of the fast changing pulse response is 0.001 and it is 4 times larger than that of the peak of the low frequency sinusoidal input. While for the $y$ magnetic field response, the pulse peak output is very small and the low changing sinusoidal magnetic field is nearly totally suppressed. Even though the $y$ direction owns to a better low frequency magnetic field suppression ability, the response of the brain magnetic field is also greatly suppressed. Thus, we can use the $x$ channel response for the brain magnetic field probing.\\
We also measured the sensitivity of the magnetometer and Figure.\ref{Figure.6} shows the result. We measured the sensitivity by the typical noise spectral method\cite{yao2016}. We let the magnetometer stayed still and the residual magnetic fields in three directions are zeroed. The output signal is recorded by a data acquire system. Note that the output signal is voltage signal. We need to change the output signal into the magnetic field signal through the scale factor. The scale factor is measured based on this method\cite{fang2016low,yao2016}. The power spectral density is calculated based on the time zone magnetic field signal. In order to clearly see the noise floor, we averaged the power spectral density in very 0.1Hz. That means in very 0.1Hz bin the average amplitude noise represents the noise in the frequency range. Note that the frequency responses of the magnetometer to different frequency magnetic field are varied. Thus the scale factors are also frequency dependent. For example, the magnetometer is not sensitive to low frequency noise, thus the scale factors in low frequency area is small while at larger frequencies, the scale factors are larger. The scale factors for the $x$ and $y$ directions are also different. It is obvious that our magnetometer is less sensitive to magnetic field in the $y$ direction. Thus the $y$ direction owns a worse sensitivity. In Figure.\ref{Figure.6}, the solid line shows the magnetic field sensitivity in the $x$ direction. There are several noise source could affect the sensitivity of the magnetometer. We see peaks at frequency range of 1-5Hz, we have figured out that the noise is from the vibration of our optical board which is floated by a mechanical vibration isolation system which can cause rotation vibration at that frequency range. We have point out that our magnetometer is sensitive to angular velocity. At lower frequency range, the 1/f noise dominant the noise band. At the frequency range of 5-10Hz, the sensitivity of 3.2fT/Hz$^{1/2}$ has been achieved. For frequency larger than 20Hz, the noise source is clearly goes down with frequency increasing. That is because we have used the low pass filter whose 3dB bandwidth is 20Hz to limit our magnetometer to 20Hz. The reason why we do this is because our magnetometer has a limited bandwidth and the auditory evoked brain magnetic field is also in this frequency range.\\
For comparison, we also show the dotted line which illustrates the sensitivity of the magnetometer if there is no low frequency magnetic field noise compensation in the $x$ direction. If there is no magnetic field noise compensation, the noise power is obvious larger at frequency smaller than 1Hz. This obvious show that the low frequency magnetic field noise compensation effect works. We also measure the sensitivity of the magnetometer in the $y$ direction. As we have shown that the magnetometer is less sensitive to magnetic field in the $y$ direction. If we see the noise spectral under 1Hz, the $y$ direction noise still goes down while that of the $x$ direction is nearly flat. Under low frequencies, the noise in the $y$ direction is smaller than that of the $x$ direction. This means that the $y$ direction owns a better magnetic field compensation effect.\\
It is reported that the magnetic field noise from the magnetic field shielding material could dominant the noise source\cite{Kornack2007,dang2010ultrahigh}. Thus we also do a calculation of our ferrite magnetic field noise. Based on the theory\cite{Kornack2007}, the thermal noise of the domain in the ferrite could produce low frequency magnetic field noise. At higher frequency the electrons in the ferrite move randomly could also produce magnetic field noise. In Figure.\ref{Figure.6}, the dash-dotted line shows the magnetic field noise originated from the 10cm inner diameter, 1cm thick and 30cm in length ferrite. This noise level is obvious smaller than the magnetometer's sensitivity. We believed that the laser power noise may be the main noise source.\\
\section{Discussion}
\label{sec:discussion}
Rotation could affect the sensitivity of the magnetometer. This is because that the nuclear spins response to rotation very obviously. Thus if we want to record moving MEGs, we need to avoid rotating moving. In Figure.\ref{Figure.5}, we only do simulation of the magnetometer. We are now trying to build a magnetic field shielding room for real measurement of the moving MEGs. This paper only gives the magnetic field compensation ability and the sensitive of the magnetometer.  We also think that better magnetic field compensation could be achieved by optimizing parameters such as the electron spins' magnetic field and the nuclear spins' magnetic field. The nuclear spins utilized in this paper is $^{21}$Ne. We need 5 hours to polarize there spins. This time is so long that it could limit the robustness of the magnetometer. Thus, other nuclear spins such as $^{129}$Xe and $^{131}$Xe should be studied.
For comparison, the method utilized in this reference\cite{boto2018} which aims to do moving MEGs measurement is based on a coil system compensation. It is obvious the magnetic field gradients could affect the compensation because only 1 magnetometer is utilized as the feedback of coil compensation. Thus there is a complicate gradient compensation system which is  utilized as well as only $40cm\times40cm\times40cm$ area is the measurement area. However, in our magnetometer, the nuclear spins in each magnetometer could compensate the background magnetic field automatically and it is in-situ magnetic field compensation. The magnetic field gradients are not need to be compensated and our measurement area will be larger.\\
\section{Conclusion}
In conclusion, we have developed an atomic magnetometer which could automatically compensate the background fluctuation magnetic field in situ. In the magnetometer, isotope enriched $^{21}$Ne atoms are utilized and they are hyper-polarized through spin exchange optical pumping. The polarized nuclear spins will automatically tracing the direction of the outside fluctuating magnetic field as well as produce an opposite direction magnetic field which will cancel the fluctuating magnetic field. This will shield the alkali atoms' spin which is utilized for brain magnetic field sensing from the fluctuating magnetic field. Due to the bandwidth of the shielding effect, the higher frequency brain magnetic field will be sensed by the atomic magnetometer while the low frequency fluctuating background magnetic field will be cancelled. \\
We have developed theory to show the ability of the shielding effect and studied its relationship with $B^e$ and $B^n$. We also studied the compensation effect experimentally. The simulation of the magnetometer's response to brain magnetic field as well as slow changing background magnetic field is also studied. Finally we measured the sensitivity of the magnetometer and 3.2fT/Hz$^{1/2}$ has been achieved in the $x$ direction.\\
\section{Declaration of Competing Interest}
The authors declare no conflict of interest.
\section{Credit authorship contribution statement}
Yao Chen designs and constructs the experiment. Some of the datas are taking by Yintao Ma and Yanbin Wang. Mingzhi Yu helps to Construct some parts of the experiment. Yao Chen also wrights the manuscript and does simulation of the magnetometer. Professor Zhuangde Jiang and Libo Zhao supervise the experiment.
\section{Acknowledgement}
This work is supported by Open Research Projects of Zhejiang Lab under grant number 2019MB0AB02, China Postdoctoral Science Foundation under grant number 2020M683462, National Natural Science Foundation of China under grant number 62103324 and Natural Science Foundation of Jiangsu under grant number BK20200244.
\bibliographystyle{unsrtnat}
\bibliography{references}  %%% Uncomment this line and comment out the ``thebibliography'' section below to use the external .bib file (using bibtex) .





% \end{thebibliography}

\end{document}